Thermal resistance by transition between collective and non-collective phonon flows in graphitic materials

[1,2]Sangyeop Lee*, [1]Xun Li, [1]Ruiqiang Guo

[1]Department of Mechanical Engineering and Materials Science, University of Pittsburgh, Pennsylvania, 15261, USA
[2]Department of Physics and Astronomy, University of Pittsburgh, Pennsylvania, 15261, USA

*sylee@pitt.edu

**ABSTRACT**

Phonons in graphitic materials exhibit strong normal scattering (N-scattering) compared to umklapp scattering (U-scattering). The strong N-scattering cause collective phonon flow, unlike the relatively common cases where U-scattering is dominant. If graphitic materials have finite size and contact with hot and cold reservoirs emitting phonons with non-collective distribution, N-scattering change the non-collective phonon flow to the collective phonon flow near the interface between graphitic material and a heat reservoir. We study the thermal resistance by N-scattering during the transition between non-collective and collective phonon flows. Our Monte Carlo solution of Peierls-Boltzmann transport equation shows that the N-scattering in graphitic materials reduce heat flux from the ballistic case by around 15%, 30%, and 40% at 100, 200, and 300 K, respectively. This is significantly larger than ~ 5% reduction of Debye crystal with similar Debye temperature (~ 2300 K). We associate the large reduction of heat flux by N-scattering with the non-linear dispersion and multiple phonon branches with different group velocities of graphitic materials.

## I. Introduction

Phonon transport in crystalline materials has been often discussed between ballistic and diffusive limits depending on sample size. The ballistic regime occurs when sample size is much smaller than the mean free path of internal phonon scattering such that the internal phonon scattering can be ignored. The diffusive regime occurs when umklapp scattering (U-scattering), which do not conserve phonon crystal momentum, is the most dominant scattering mechanism. For the diffusive regime, sample size should be larger than the mean free paths (MFPs) of U-scattering. There is another regime of phonon transport, called hydrodynamic regime, which rarely occur compared to the ballistic and diffusive regimes. The hydrodynamic phonon transport occurs when most internal phonon scattering processes are a momentum conserving type (i.e., normal scattering and hereafter N-scattering) and thus do not directly cause thermal resistance. The hydrodynamic regime was predicted and experimentally observed several decades ago [1-4].

The hydrodynamic regime recently received a renewed attention after first-principles-based calculations predicted its significance in graphitic materials including graphene [5, 6], single wall carbon nanotubes (SWCNTs) [7], and graphite [8]. These graphitic materials commonly share flexural phonon modes and large Debye temperature, both of which together lead to the significant hydrodynamic phonon transport [5]. The first-principles-based calculations show that the MFPs of N-scattering are several orders of magnitude smaller than those of U-scattering and thus N-scattering is the dominant scattering mechanism [9].

The N-scattering does not directly cause thermal resistance because of its momentum conserving nature. In particular, when N-scattering is the only scattering mechanism and a sample is infinitely large, N-scattering does not cause any thermal resistance. After the sufficient number of N-scattering events, phonon system has the displaced Bose-Einstein distribution defined as

$$f^{\text{disp}} = \left( \exp\left( \frac{\hbar(\omega - \mathbf{q} \cdot \mathbf{u})}{k_B T} \right) - 1 \right)^{-1} \qquad (1)$$

where $\mathbf{q}$ and $\mathbf{u}$ are a phonon wavevector and a displacement (or drift velocity). Once the displaced Bose-Einstein distribution is established, N-scattering does not further alter the phonon distribution and thus does not cause any thermal resistance. The phonons with the displaced distribution function can continue to flow even without any temperature gradient, resulting in the infinite thermal conductivity [10].

The N-scattering, however, affects thermal resistance in more realistic cases where the U-scattering exist or sample size is finite. When both N- and U-scattering exist and a sample is still infinitely large, N-scattering, combined with U-scattering, can cause thermal resistance. Phonon states with a small wavevector usually have very weak U-scattering, but N-scattering can bring the energy in the small wavevector states to large wavevector states where U-scattering is relatively strong. Thus, the N-scattering combined with U-scattering cause thermal resistance. However, if sample size is finite and smaller than the mean free path of U-scattering, diffuse boundary scattering plays a more important role for thermal resistance than U-scattering. Particularly when a sample has infinite length but finite width and phonon flows along the length direction, a major resistance mechanism is phonon viscous damping similar to the molecular Poiseuille flow case. In such a case, stronger N-scattering leads to less phonon viscosity and smaller thermal resistance. This leads to the thermal conductivity values that increase superlinearly with width and increase faster than the ballistic conductance with temperature [11, 12] which was used to experimentally verify the phonon Poiseuille flow [3].

Our study focuses on another case where a sample has finite length but infinite width and N-scattering is much stronger than U-scattering. This case is relevant to practical applications where thin-film of ultrahigh thermal conductivity materials are used for thermal management. If the sample thickness lies in the gap between the MFPs of N- and U-scatterings, actual phonon transport can be roughly approximated as the N-scattering only case. For the sake of simplicity, here we assume that our sample is placed between two black bodies that emit phonons with a stationary Bose-Einstein distribution (i.e., the Bose-Einstein distribution with zero drift velocity) and do not reflect incoming phonons. The emitted phonons would experience N-scattering and their distribution accordingly changes from the stationary Bose-Einstein distribution (i.e., non-collective phonon flow) to the displaced Bose-Einstein distribution (i.e., collective phonon flow). After the displaced Bose-Einstein distribution is established, the N-scattering does not further alter the distribution function and thus does not cause thermal resistance like the aforementioned case of the infinite thermal conductivity [13]. However, thermal resistance can occur during the transition between non-collective and collective flows near the boundaries.

In this paper, we solve the Peierls-Boltzmann transport equation (PBE) using a Monte Carlo (MC) method to calculate the thermal resistance during the non-collective to collective transition of phonon flow by N-scattering in graphitic materials including (10,10) SWCNT, (20,20)

SWCNT, graphene, and graphite. The thermal resistance of graphitic materials is compared to that of a Debye crystal with similar Debye temperature (2300 K).

## II. Methods and Approaches

The transition from non-collective to collective phonon flows can be correctly captured only when the PBE is solved in both real and reciprocal spaces. The PBE for a finite size sample has been often solved in a reciprocal space only [9, 14, 15] for calculating thermal conductivity values. In this case, the distribution function in a real space is assumed to follow the stationary Bose-Einstein distribution with a constant temperature gradient. The finite sample size effect was considered with a simple expression of boundary scattering rate, $\tau_{B,i} = L(2|\mathbf{v}_i|)^{-1}$, where $\tau_{B,i}$ is the boundary scattering rate of phonon state $i$ and $L$ is the characteristic size of sample. This simple approach works well for the cases where sample is much larger than the relaxation length of heat flux (e.g., MFPs of U-scattering) so that the distribution function can be safely assumed to be homogeneous in a real space. In our case, however, the spatially homogeneous distribution function cannot be assumed because the transition between non-collective and collective phonon flows occur near boundaries and the phonon distribution function accordingly changes in a real space.

We solve the following energy-based PBE in both real and reciprocal spaces by employing a deviational MC method [16]:

$$\mathbf{v}_i \cdot \nabla \left( \omega_i f_i^d \right) = \left( \frac{d\left( \omega_i f_i^d \right)}{dt} \right)_{\text{scatt}} \tag{2}$$

where $f_i^d$ is a deviational distribution function defined as the difference between an actual phonon distribution ($f_i$) and the stationary Bose-Einstein distribution ($f_i^0$) at a global equilibrium temperature ($T_0$):

$$f_i^d = f_i - f_i^0(T_0) \tag{3}$$

Hereafter $f_i^0$ represents $f_i^0(T_0)$ unless temperature is specified. The sample and boundary conditions are schematically shown in Fig. 1. The energy-based PBE is chosen over the regular PBE because the energy conservation can be strictly satisfied during scattering processes in the energy-based PBE. In our MC simulation, all sample particles (hereafter particles) are set to carry either positive or negative unit energy. A positive particle which carries a positive unit energy contributes to the positive $\omega_i f_i^d$. A negative particle which carries a negative unit energy contributes to the negative $\omega_i f_i^d$. Therefore, by simply conserving the net number of particles during scattering, the total energy conservation is strictly satisfied. The MC simulation consists of two steps during a given time step: advection and scattering. The advection step allows particles to fly with their group velocities for a given amount of time duration. The scattering step stochastically determines the occurrence of scattering.

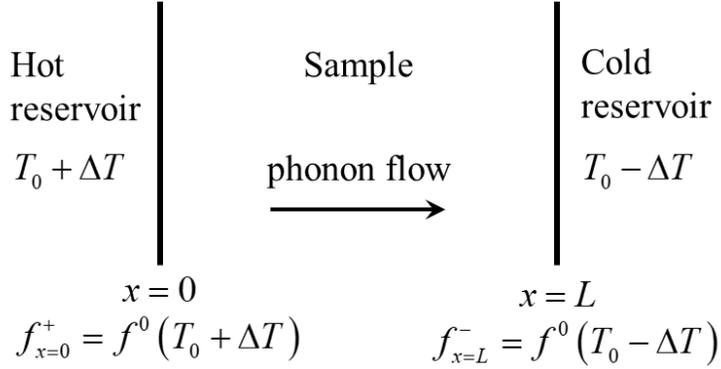

**Fig. 1.** A schematic of a sample with boundary conditions. The superscripts + and − represent positive and negative group velocities along the *x*-direction.

We use two different scattering models in our study; one is the Callaway's scattering model with a constant scattering rate and the other is full three-phonon scattering matrix from *ab initio* calculation. The Callaway's scattering model was chosen to study the effects of phonon dispersion on the thermal resistance for various graphitic materials since the thermal resistance in our case depends on the phonon dispersion as will be shown later. To our best knowledge, there is no previous study using the MC method to solve the PBE with the Callaway's scattering model and here we briefly introduce our MC method for this scattering model. The details of MC method employing full scattering matrix can be found somewhere else [12, 17].

The energy-based PBE employing the Callaway's scattering model with N-scattering is

$$\mathbf{v}_i \cdot \nabla \omega_i f_i^d = -\omega_i \frac{f_i^d - f_i^{d,disp}}{\tau} \qquad (4)$$

where $\tau^{-1}$ is the rate of N-scattering, which is assumed a constant for all phonon modes in this study. The $f_i^{d,disp}$ is the deviation of the displaced Bose-Einstein distribution from the stationary Bose-Einstein distribution at the global equilibrium temperature:

$$f_i^{d,disp}(T(x),u(x)) = f_i^{disp}(T(x),u(x)) - f^0 \tag{5}$$

where $T(x)$ and $u(x)$ are local equilibrium temperature and displacement at position $x$, respectively. For general cases where both N- and U-scattering exist, the algorithm presented here for N-scattering only case can be easily combined with that of U-scattering case from the literature [18, 19]. Assuming the small deviation from the equilibrium distribution, i.e., $q_{x,i}u(x) \ll \omega_i$ and $|T(x)-T_0| \ll T_0$, the displaced Bose-Einstein distribution can be linearized and the Eq. (4) can be expressed as

$$\mathbf{v}_i \cdot \nabla \omega_i f_i^d = -\frac{1}{\tau}\left(\omega_i f_i^d - \left(\omega_i f_i^{d,disp,even} + \omega_i f_i^{d,disp,odd}\right)\right) \tag{6}$$

where

$$\omega_i f_i^{d,disp,even} = \frac{\hbar \omega_i^2}{k_B T_0} f^0 (f^0 + 1)\left(\frac{T(x)-T_0}{T_0}\right) \tag{7}$$

$$\omega_i f_i^{d,disp,odd} = \frac{\hbar \omega_i q_{x,i}}{k_B T_0} f^0 (f^0 + 1) u(x) \tag{8}$$

The $\omega_i f_i^{d,disp,even}$ is an even function with respect to $q_x$ representing the deviation of local energy density from the global equilibrium case. The $\omega_i f_i^{d,disp,odd}$ is an odd function with respect to $q_x$ and associated with a net energy flow.

As the Callaway's scattering model requires a local equilibrium temperature and a local displacement, we divide the real space domain into many small control volumes. Within each control volume, temperature and displacement are assumed to be spatially homogeneous. All scattering processes should be simulated such that total energy and momentum are conserved within a control volume. Below we present the MC algorithm for the Callaway's scattering model.

i) The scattering of each particle is determined by comparing a random number between 0 and 1 to the probability of scattering, $P = 1 - \exp(-\Delta t/\tau)$, where $\Delta t$ is a given duration of a time step. If the random number is smaller than the probability of scattering, a particle is determined to be scattered and removed. The total energy and *x*-direction momentum that the scattered particles carry are counted for enforcing energy and momentum conservation later.

ii) After the occurrence of scattering for all particles in a control volume is determined, new particles are generated according to the local equilibrium distribution function, $\omega_i f_i^{d,disp,even} + \omega_i f_i^{d,disp,odd}$. For the even part of local equilibrium distribution, $\omega f^{d,disp,even}$, particles are drawn with the normalized distribution, $\omega_i^2 f_i^0 (f_i^0 + 1) / \sum_i \omega_i^2 f_i^0 (f_i^0 + 1)$. The number of particles and their sign of energy are determined such that the total energy of generated particles is the same as that of scattered particles. For the odd part of local equilibrium distribution, $\omega f^{d,disp,odd}$, we generate pairs of particles consisting of a positive and a negative energy particles. The positive particle is drawn from $q_x > 0$ with the normalized distribution, $\omega_i |q_{x,i}| f_i^0 (f_i^0 + 1) / \sum_i \omega_i |q_{x,i}| f_i^0 (f_i^0 + 1)$. The negative particle is drawn from $q_x < 0$ with the same normalized distribution. More pairs are generated until the total momentum of generated particles equals to that of scattered particles. Each pair's net energy is zero and thus generating pairs do not affect the local energy density.

The developed MC code with the Callaway's scattering model was validated against the previously reported semi-analytic solutions of the PBE in a wide range of Knudsen number [20, 21]. The semi-analytic solution assumes a sample with infinite length but finite width. In Fig. 2, we show the comparison between our MC result and the semi-analytic solution of the PBE with the Callaway's scattering model when there is no U-scattering [21]. Our MC solution shows a good agreement with the semi-analytic solution, verifying our MC method and code in a wide range of Knudsen number. Both MC and semi-analytic solutions show the superlinear dependence of thermal conductivity with width, which is a feature of the phonon Poiseuille flow [11, 12]. For a sample with finite length and infinite width, our MC results also agree well with the data from a previous work based on the semi-analytic solution of PBE [22] as shown in the inset of Fig. 4(e).

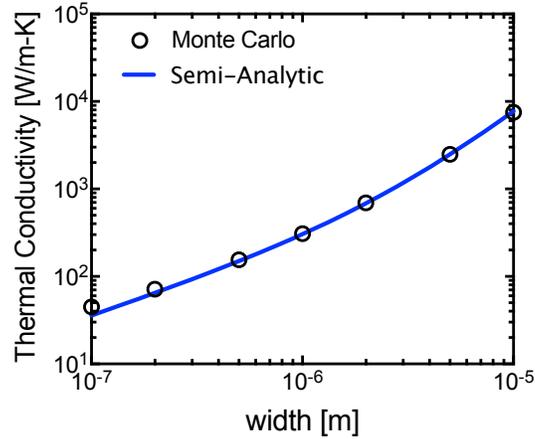

**Fig. 2.** Comparison between Monte Carlo (MC) and Semi-Analytic (SA) solutions of the PBE [21]. The Debye model is used with a group velocity of $10^4$ m/s and a Debye temperature of 1200 K. The N-scattering rate is fixed at $10^{10}$ s$^{-1}$ for the MC and SA data. Temperature is assumed to be 100 K.

The developed MC code is used with realistic phonon dispersions of SWCNTs, graphene, and graphite. The phonon dispersion of (10,10) and (20,20) SWCNTs are from a previous work using an optimized Tersoff potential [23]. For graphene and graphite, the phonon dispersion was calculated using the density functional theory calculation. The second order force constants were calculated using the Vienna Ab initio Simulation Package (VASP) [24] with the projector-augmented-wave pseudopotentials [25] and local density approximation for exchange-correlation energy functional [26]. For graphite, the van der Waals interaction is included with a non-local correlation functional, optB88-vdW. Then the phonon dispersion was calculated the PHONOPY package [27].

## III. Results and Discussion
### III.1. SWCNTs, graphene, and graphite with a constant N-scattering rate

In Fig. 3, we present the distribution of temperature and drift velocity in a (20,20) SWCNT at different Knudsen numbers. We assume the Callaway's scattering model with a constant N-scattering rate. The Knudsen number is defined as $\tau_N |\mathbf{v}|/L$ where $\mathbf{v}$ is the largest group velocity of acoustic modes. In Fig. 3(a), near the hot reservoir boundary, the drift velocity increases while temperature decreases, indicating that the phonons form the collective flow through N-scattering processes with the expense of a temperature drop. For larger sample length shown in Fig. 3(b) and (c), the drift velocity and temperature are further changed as a distance from the boundary increases, but become constants when the distance is around 3 μm, which is comparable to the MFP. This is because the phonon distribution at 3 μm becomes the same as the displaced distribution function. After the displaced distribution function is established, N-scattering does not further change the distribution function. As a result, the phonons can continue to flow even without a temperature drop, causing the infinite local thermal conductivity Peierls suggested. A similar behavior was reported using the first-order solution of PBE [13] and using the semi-analytic solution of PBE for Debye model [22]. The distribution and temperature change once again near the cold reservoir boundary because of the interaction with phonons emitted from the cold boundary with a non-collective distribution.

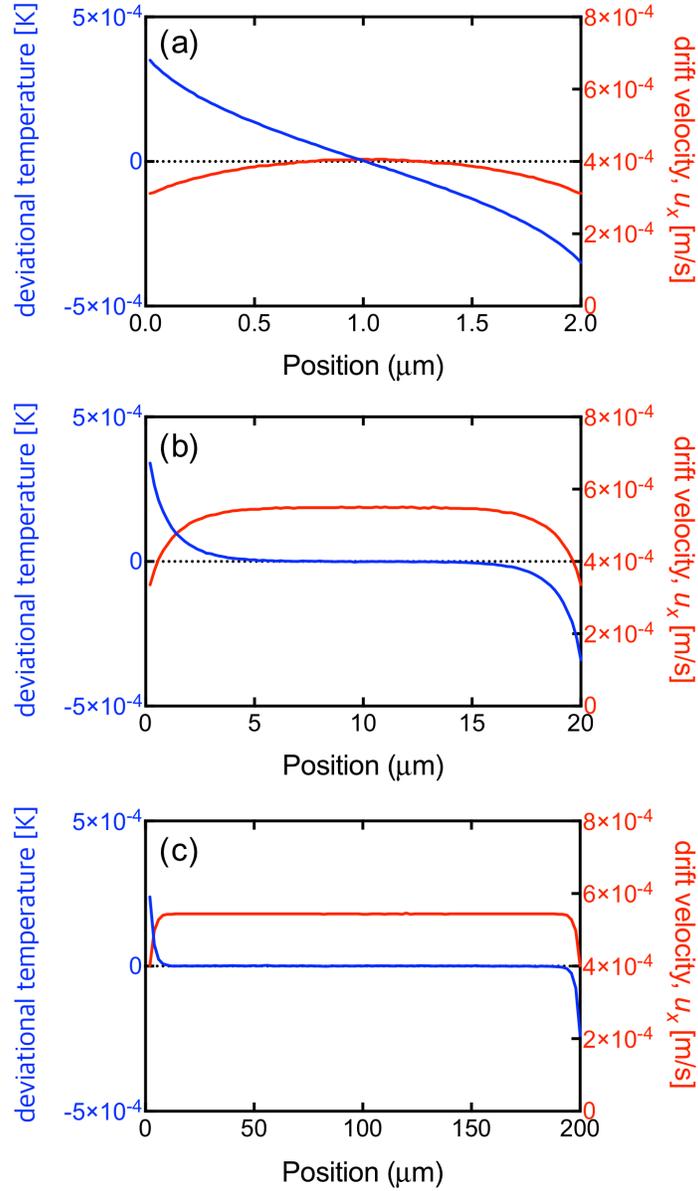

**Fig. 3.** The profile of temperature and drift velocity along (20,20) SWCNT with length of (a) 2 μm, (b) 20 μm, and (c) 200 μm, which represent the Knudsen number of 1.0, 0.1, and 0.01, respectively. The left *y*-axis is the deviational temperature defined as $T(x)-T_0$ where $T_0$ is 300 K. The deviational temperatures of hot and cold reservoirs are 0.001 and -0.001 K, respectively. The right *y*-axis is the drift velocity.

The transition between collective and non-collective phonon flows cause thermal resistance as we can see the temperature drop near the boundaries in Fig. 3. The thermal resistance due to this transition is shown in Fig. 4. The *y*-axis is the ratio between the heat flux under N-scattering ($q_\text{H}''$) and the heat flux of purely ballistic case ($q_\text{B}''$). The ratio, $q_\text{H}''/q_\text{B}''$, measures the resistance caused by N-scattering. For all graphitic materials and Knudsen numbers, the heat flux $q_\text{H}''$ is smaller than $q_\text{B}''$, indicating that N-scattering causes thermal resistance even without U-scattering. The heat flux $q_\text{H}''$ decreases with an increasing inverse of Knudsen number (i.e., length of a sample) until the inverse Knudsen number becomes around unity. Then, the heat flux $q_\text{H}''$ does not further decrease with the inverse Knudsen number. As discussed above, the thermal resistance occurs near the boundary within a distance comparable to MFP and the N-scattering occuring far from the boundaries do not contribute to thermal resistance. As a result, the thermal resistance in the ideal hydrodynamic regime where N-scattering is the only scattering mechanism exhibits a very different behavior from ballistic and diffusive regimes. The thermal resistance in ballistic regime is a constant regardless of length and that in diffusive regime linearly increases with sample length. The thermal resistance in hydrodynamic regime increase with length when the length is smaller than MFP of N-scattering and then does not further increase with length when the length is larger than MFP of N-scattering.

We show the resistance by N-scattering for Debye model and graphitic materials in Fig. 4. While the $q_\text{H}''$ is around 95% of $q_\text{B}''$ for Debye model, which agree well with the semi-analytic solution reported in Ref. [22], the $q_\text{H}''$ is around 85%, 70%, and 60% of the ballistic heat flux $q_\text{B}''$ at 100, 200, and 300 K, respectively, for graphitic materials. Therefore, even when N-scattering is the only internal scattering mechanism, it can cause significant thermal resistance near the boundary between a graphitic sample and a reservoir. It is also noteworthy that the ratio $q_\text{H}''/q_\text{B}''$ is a constant regardless of temperature for the Debye crystal but decreases with temperature for graphitic materials.

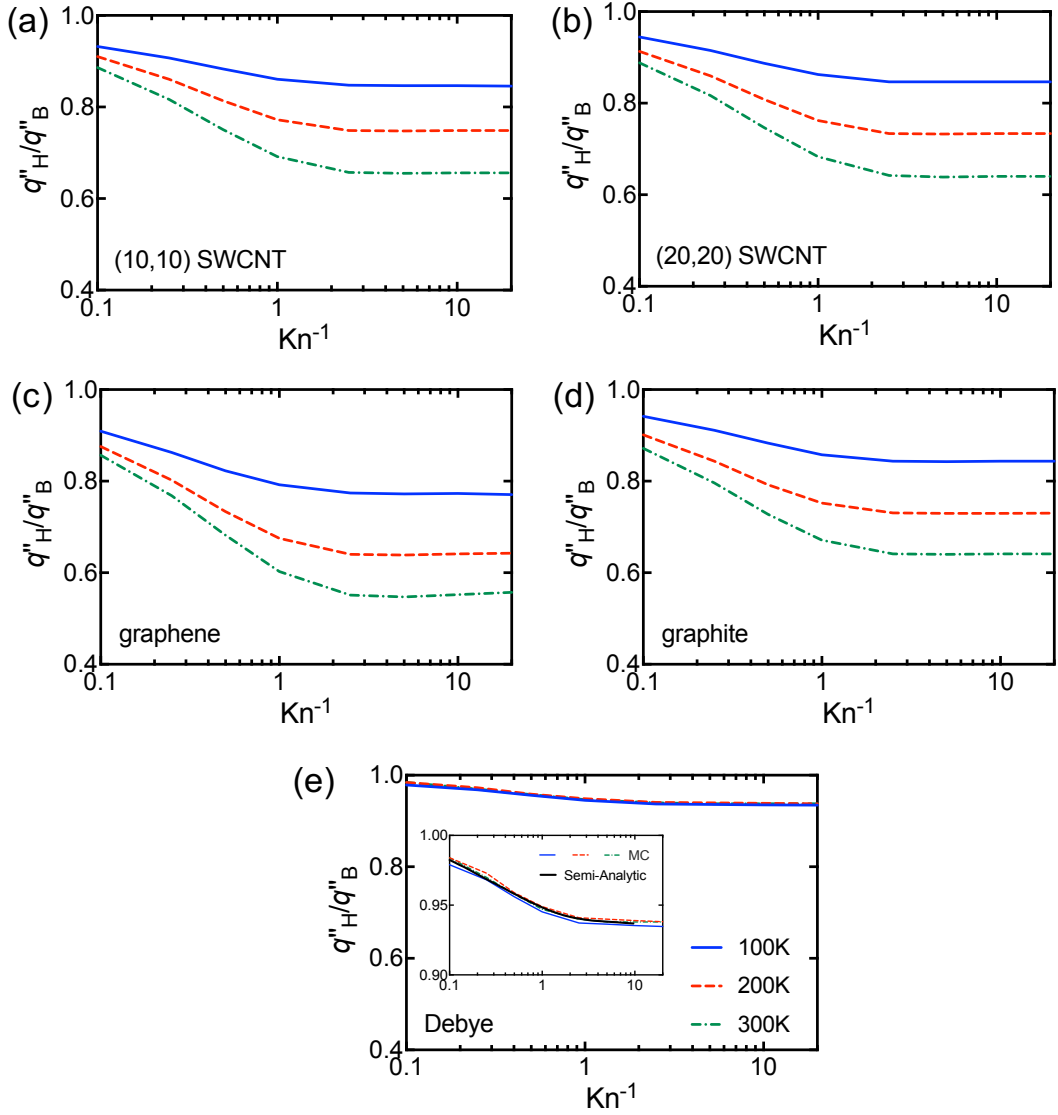

**Fig. 4.** Heat flux under N-scattering ($q''_H$) compared to the heat flux of purely ballistic case ($q''_B$) for (a-d) graphitic materials and (e) Debye model showing the reduction of heat flux due to N-scattering. For (e), the Debye temperature is assumed as 2300 K similar to the graphitic materials. The inset of (e) shows a good agreement between our MC result and data from the literature based on the semi-analytic solution of PBE [22], verifying our MC method.

To explain the large thermal resistance by N-scattering in graphitic materials, we consider the entropy generation upon N-scattering. For the sake of simplicity, we ignore the spatial variation of $f_i$. Instead, we simply assume a process where all phonon particles emitted from hot and cold reservoirs experience N-scattering in the middle of sample. The rate of entropy generation due to scattering is [28]

$$\dot{S}_{scatt} = \frac{1}{TN_qV} \sum_i \phi_i \dot{f}_{scatt,i} \qquad (9)$$

where $\phi_i$ represents the deviation of distribution function from the stationary Bose-Einstein distribution, which is defined as

$$f_i = f_i^0 - \phi_i \frac{\partial f_i^0}{\partial(\hbar\omega_i)} \qquad (10)$$

Assuming the Callaway's scattering model with a constant N-scattering rate, Eq. (9) can be written as

$$\dot{S}_{scatt} = \frac{1}{TN_qV\tau_N} \sum_i \phi_i \left(f_i - f_i^{disp}\right) \qquad (11)$$

The phonon distribution before scattering follows the stationary Bose-Einstein distribution of hot and cold reservoirs where the phonon particles are emitted from:

$$f_i = f_i^0 + \frac{\partial f_i^0}{\partial T}\Delta T \quad \text{for } v_{x,i} > 0 \qquad (12)$$

$$f_i = f_i^0 - \frac{\partial f_i^0}{\partial T}\Delta T \quad \text{for } v_{x,i} < 0 \qquad (13)$$

Then, Eq. (11) can be written as

$$\dot{S}_{scatt} = \left(\frac{\Delta T}{T}\right)^2 \frac{\hbar^2}{\tau_N k_B T^2 N_q V} \sum_i f_i^0\left(f_i^0+1\right)\omega_i |q_{x,i}|\left(v_{x,i}^* - u_x'\right) \qquad (14)$$

where $v_{x,i}^*$ is $\omega_i/|q_{x,i}|$ and $u_x'$ is the drift velocity per temperature difference defined as

$$u_x' = u_x \left(\Delta T/T\right)^{-1} \qquad (15)$$

The $u_x'$ can be derived from the momentum conservation as:

$$u_x' = \frac{\sum_i |q_{x,i}|\omega_i f_i^0\left(f_i^0+1\right)}{\sum_i q_{x,i}^2 f_i^0\left(f_i^0+1\right)} \qquad (16)$$

The Eq. (14) confirms that the entropy generation (or thermal resistance) by N-scattering is determined by phonon dispersion if N-scattering is strong enough to establish the displaced

Bose-Einstein distribution. In Eq. (14), the term $v_{x,i}^*$ is from the phonon distribution of emitted particles in Eqs. (12) and (13). This indicates that the temperature difference, $\pm\Delta T$, drives the phonon flow with the displacement of $v_{x,i}^*$, which may vary depending on phonon modes. The N-scattering change the displacement $v_{x,i}^*$ to $u_x'$ which is a constant for all phonon modes. Then, the magnitude of entropy generation is proportional to the difference between $v_{x,i}^*$ and $u_x'$. If $v_{x,i}^*$ is a constant (e.g., a Debye phonon dispersion in 1D space), $u_x'$ is the same as $v_{x,i}^*$ and the entropy generation should be zero which is the lower limit of entropy generation by the Boltzmann's *H*-theorem. We confirmed that $q_H''/q_B''$ is unity for this case using the MC simulation. However, if $v_{x,i}^*$ largely varies with phonon states, the entropy generation can become large. For the Debye phonon dispersion in 2D and 3D materials, $v_{x,i}^*$ varies with the phonon propagation direction, resulting in non-zero resistance upon N-scattering as shown in Fig. 4e. The ratio $q_H''/q_B''$ for Debye model does not change with temperature as the variance of $v_{x,i}^*$ is associated with the phonon propagation direction only. For graphitic materials, $v_{x,i}^*$ varies more significantly compared to the Debye model due to the highly non-linear phonon dispersion and many phonon branches, causing the large thermal resistance upon N-scattering in Fig. 4a-d. Also, as temperature increases, the variance of $v_{x,i}^*$ becomes larger as phonon states at a high frequency have much different group velocities from phonon states at a lower frequency, which result in larger resistance by N-scattering at higher temperature.

### III.2. Graphene using a full three-phonon scattering matrix

We have discussed the thermal resistance due to the transition between non-collective and collective phonon flows using the PBE with the Callaway's scattering model assuming a constant N-scattering rate and no U-scattering. However, real graphitic materials have a wide range of MFPs of both N- and U-scattering processes. We use the MC solution of PBE with *ab initio* three-phonon scattering matrix to discuss how much portion the hydrodynamic thermal resistance near the boundaries contribute to the total resistance.

In Fig. 5, we compare the ratio of $q''_H$ and $q''_B$ in suspended graphene. We particularly compare the ratios $(q''_H/q''_B)_{N+U}$ and $(q''_H/q''_B)_N$. The former is the ratio when both N- and U-scatterings are included and the latter is the ratio when U-scattering is removed and N-scattering is the only internal scattering mechanism. The difference between $(q''_H/q''_B)_{N+U}$ and $(q''_H/q''_B)_N$ gives a rough estimate of how much N-scattering contritube to the total resistance. At 100 K in Fig. 5a, when length is below 10 μm, $(q''_H/q''_B)_{N+U}$ and $(q''_H/q''_B)_N$ are almost identical, reaching 0.8 at the length of 10 μm. Therefore, for the length below 10 μm at 100 K, N-scttering near the boundaries can reduce the heat flux (or thermal conductivity) from the ballistic case by 20 %. When length is above 10 μm, $(q''_H/q''_B)_{N+U}$ sharply decrease with length due to U-scattering while $(q''_H/q''_B)_N$ remains the same. It is noteworthy that the converged value of $(q''_H/q''_B)_N$, which is around 0.8, agree well with $q''_H/q''_B$ from a simple scattering model in Fig. 4c. This confirms that the hydrodynamic resistance near the boundaries is determined by the shape of phonon dispersion not actual N-scattering rate, if the sample length is larger than the MFP of N-scattering.

At 300 K in Fig. 5b, although U-scattering rate is increased from 100 K case, the thermal resistance by N-scattering still contributes a significant portion to total resistance for length smaller than 1 μm. The $(q''_H/q''_B)_N$ decrease with sample length and reach around 0.5 when the length is 1 μm, meaning that the heat flux (or conductivity) is reduced by 50 % from the ballistic case by N-scattering. The $(q''_H/q''_B)_N$ converging to 0.5 at 300 K agrees well with the result from the simple scattering model shown in Fig. 4c. However, when sample length is longer than 1 μm, U-scattering significantly reduce the heat flux and the thermal resistance by N-scattering is relatively not important.

It is noteworthy that a recent study on four-phonon scattering suggests that the four-phonon scattering plays an important role in suspended graphene with N-scattering being much stronger than U-scattering like three-phonon scattering [29]. Thus, we expect that the hydrodynamic regime is still significant even with four-phonon scattering. The inclusion of four-phonon scattering would not change the converged value of $(q''_H/q''_B)_N$, because this ratio is determined by the phonon dispersion not by scattering rates as long as the sample is longer than the MFP of N-scattering.

However, the length at which $(q''_H/q''_B)_N$ is converged will be reduced, as the inclusion of four-phonon scattering will reduce the MFP of N-scattering.

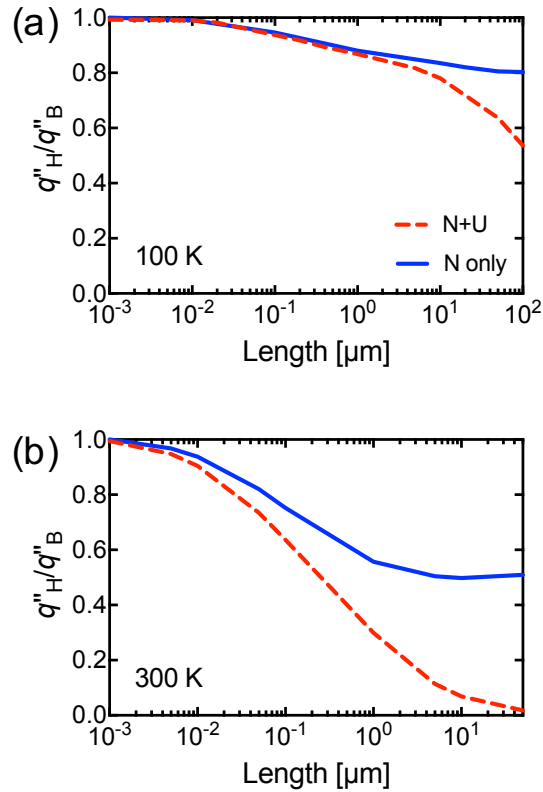

**Fig. 5.** Heat flux under N-scattering ($q''_H$) compared to the heat flux of purely ballistic case ($q''_B$) in graphene from the PBE with a full three-phonon scattering matrix. The solid line represents N-scattering only case where U-scattering is removed and the dashed line represents for both N- and U-scattering processes.

## IV. Conclusion

We have showed that large thermal resistance can be caused by N-scattering even without U-scattering in graphitic materials with finite length. The thermal resistance by N-scattering in finite length sample occurs when non-collective flow of phonons emitted from a heat reservoir is changed to collective flow by N-scattering. Assuming that the sample length is larger than MFP of N-scattering, the resistance due to N-scattering is determined by the shape of phonon dispersion, i.e., the variance of $\omega_i/|q_{x,i}|$. This is because the finite temperature difference of heat reservoirs drive phonon flow with a displacement of $\omega_i/|q_{x,i}|$ while N-scattering relax this displacement to $u'_x$ which is a constant for all phonon modes. Therefore, the entropy generation upon N-scattering is proportional to the difference of $\omega_i/|q_{x,i}|$ and $u'_x$. The N-scattering do not cause any resistance for 1D Debye model where $\omega_i/|q_{x,i}|$ is a constant. However, for graphitic materials, $\omega_i/|q_{x,i}|$ significantly vary with phonon states due to many branches and highly non-linear phonon dispersion, resulting in the large resistance when non-collective phonon flow becomes the collective phonon flow upon N-scattering. We show that the heat flux or conductivity is reduced by around 15%, 30%, and 40% from ballistic case at 100, 200, and 300 K, respectively, in graphitic materials. High thermal conductivity materials often show strong N-scattering compared to U-scattering. Our observation shows that the shape of phonon dispersion can be one important parameter for developing high thermal conductivity materials particularly when the material is used as a thin-film where heat flows along the cross-plane direction.

**Acknowledgement**

We acknowledge support from National Science Foundation (Award No. 1705756 and 1709307). The simulation was performed using the Linux clusters of Extreme Science and Engineering Discovery Environment (XSEDE) through allocation TG-CTS180043 and Center for Research Computing at the University of Pittsburgh. S.L. thank L. Lindsay for providing the phonon dispersion of SWCNTs.